\documentstyle[prd,aps,twocolumn]{revtex}
\begin{document}

\input epsf
\draft

\title{Quantum Fluctuations and Dynamical Chaos:  An Effective Potential
Approach}

\author{Sergei G. Matinyan}

\address{Department of Physics, Duke University, Durham, NC 27708-0305
\\
Yerevan Physics Institute, Armenia}

\author{Berndt M\"uller}

\address{Department of Physics, Duke University, Durham, NC 27708-0305}

\date{\today}

\maketitle

\begin{abstract}
We discuss the intimate connection between the chaotic dynamics of a
classical field theory and the instability of the one-loop effective
action of the associated quantum field theory.  Using the example of
massless scalar electrodynamics, we show how the radiatively induced
spontaneous symmetry breaking stabilizes the vacuum state against chaos, 
and we speculate that monopole condensation can have the same effect in 
non-Abelian gauge theories.
\end{abstract}


\section{Introduction}

The effective potential $\Gamma(\phi_{\alpha})$ describes the energy
density of a quantum field theory under the constraint of the prescribed
field expectation values $\phi_{\alpha}$, assumed to be constant in
space and time.  It is well established that $\Gamma$, as the Legendre 
transform of the generating functional of connected Green functions
\begin{equation}
W[J_{\alpha}] = -i \ln\, \int [d\Phi]\; e^{-iS[\Phi_{\alpha}] -
iJ_{\alpha}\cdot \Phi_{\alpha}}, \label{e1}
\end{equation}
is always real and a convex function of the expectation values
$\phi_{\alpha}=\langle\Phi_{\alpha}\rangle$ \cite{KSy70,JI75}.  On the 
other hand, it is a familiar property of many quantum field theories that 
the one-loop effective potential $\Gamma^{(1)}$ has an imaginary part in 
regions where the tree-level potential $U(\phi_{\alpha})$ is concave.  A 
well-known example is the double-well quartic scalar potential
\begin{equation}
U(\phi) = {\lambda\over 4} (\phi^2-v^2)^2
\end{equation}
in the region $\phi^2 < v^2/3$.  The resolution of the apparent
paradox is provided by the observation that the Gaussian approximation
for the functional integral (\ref{e1}) fails when $\phi^2 <v^2$, and
hence the naive loop expansion does not apply.  A better approximation
is obtained by a superposition of two Gaussians, centered at $\Phi=\pm
v$ and weighted so that $\langle\Phi\rangle = \phi$.  This procedure
corresponds, effectively, to a Maxwell construction of the true
effective potential \cite{CT72,JSL69,LOR76,YF83,CMB83,LOR86}.  

The condition 
for the reality of the (naive) one-loop effective potential $\Gamma^{(1)}$ 
is that the matrix of second derivatives of the bare potential $U$
\begin{equation}
M_{\alpha\beta}^2 =
{\partial^2U\over\partial\phi_{\alpha}\partial\phi_{\beta}} \label{e2}
\end{equation}
does not have a negative eigenvalue.  $M_{\alpha\beta}^2$ is the
{\it mass} {\it matrix} for quantum fluctuations around the
expectation values $\phi_{\alpha}$ of the fields $\Phi_{\alpha}$.  If
the matrix $M^2$ has a negative eigenvalue, there exists a direction
in which fluctuations grow exponentially with time, indicating the
instability of the classical field configuration $\Phi_{\alpha}=
\phi_{\alpha}$.  The resulting imaginary part of the effective
potential describes the {\it decay rate} (per unit space) of the
unstable field configuration \cite{SC74,EJW87}.

The matrix of second derivatives (\ref{e2}) also plays a role in the
classical field theory as {\it stability matrix} for a classical field
trajectory against small perturbations of the initial conditions.
Here one looks at spatially homogeneous, but time dependent field
configurations $\phi_{\alpha}(t)$ which evolve according to the
classical field equations.  If the stability matrix $M^2$ has negative
eigenvalues, the trajectory $\phi_{\alpha}(t)$ is exponentially
sensitive to the initial conditions resulting, in general, in
deterministic chaotic dynamics of the fields.  Dynamical chaos of this
type is ubiquitous in field theories with more than one dynamical
field. Familiar examples are:  scalar electrodynamics, Yang-Mills
theories, and bilinearly coupled theories involving two scalar
fields.  The dynamical instability of the trajectories of classical
field configurations finds its expression in the existence of {\it
positive Lyapunov exponents} $\lambda_{\nu}>0$.  The presence of
negative eigenvalues of the stability matrix $M^2$ is a well known 
criterion (Toda-Brumer criterion \cite{MT74,PB74}) for the presence of 
the (local) dynamical instability expressed by the positive Lyapunov 
exponents.  These considerations show that the dynamical chaos of 
classical fields and the complexity of the one-loop effective potential 
of quantum fields are closely connected.

The fact that the {\it exact} effective potential $\Gamma$ is always
real indicates that the classical chaos must be suppressed in the full
quantum field dynamics.  However, this implies that the Gaussian
(loop) expansion must break down in these cases.  As a result, the
vacuum state of classically chaotic quantum fields must acquire a
nontrivial structure.  As we shall discuss below, the vacuum
instabilities can sometimes be avoided by the mechanism of spontaneous
symmetry breaking, e.g. in the case of gauge theories coupled to a
scalar (Higgs) field.  The emergence of a dynamical mass due to the
vacuum expectation value of the scalar field suppresses chaos and
makes the effective potential real in the vicinity of the vacuum
state \cite{SGM81}.  In other cases, such as Yang-Mills fields, the vacuum 
must acquire a much more complex structure, and the instabilities are
avoided by the confinement of elementary field excitations in the
infrared domain.

In the next section, we discuss dynamical chaos of classical fields in
the context of scalar electrodynamics (SED).  In section III we study
the instabilities of quantum fluctuations in SED and discuss the
stabilizing mechanism of spontaneous symmetry breaking.  Section IV is
devoted to the investigation between the imaginary part of the
effective potential in the loop expansion and dynamical chaos in SED.
Section V contains some remarks concerning the application of these
ideas to Yang-Mills theories.

\section{Classical Scalar Electrodynamics}

\subsection{Equations of Motion}

Let us first consider Classical Scalar Electrodynamics (SED) without
self-interaction of the scalar field $\phi$.  Later we will introduce 
the self-interaction of the scalar field with and without spontaneous 
symmetry breaking (SSB) but it is convenient to neglect it at first,
in order to understand the origin of dynamical chaos in SED.

The Lagrangian density for the system with bare scalar mass $m_0$ 
is given by
\begin{equation}
L = -{1\over 4}F_{\mu\nu}F^{\mu\nu} + (D_{\mu}\phi)^*
(D^{\mu}\phi) - m_0^2\phi^*\phi \label{2.1}
\end{equation}
where $F_{\mu\nu}=\partial_{\mu}A_{\nu} - \partial_{\nu}A_{\mu}$, and
$D_{\mu}= \partial_{\mu}+ie A_{\mu}$.  The corresponding equations of
motion are:
\begin{eqnarray}
(D_{\mu}D^{\mu}+m_0^2)\phi &= &0 \label{2.2a} \\
\partial_{\nu} F^{\mu\nu}&= &j^{\mu} \label{2.2b}
\end{eqnarray}
with the conserved current
\begin{eqnarray}
j_{\mu} &=& -ie(\phi^*D^{\mu}\phi - \phi D^{\mu}\phi^*) \nonumber \\
&=& -ie(\phi^*\partial^{\mu}\phi -\phi\partial^{\mu}\phi^*) +
2e^2\phi^*\phi A^{\mu}. \label{2.3}
\end{eqnarray}

In the following, we consider the case of spatially homogeneous
classical fields $A_{\mu}(t)$ and $\phi(t)$ for which the study of
chaos is extremely simplified.  The assumption of spatial homogeneity
is not Lorentz invariant, but any solution of the equations of motion 
obtained in this way can be boosted into another reference frame.  It 
is easy to see that, in the gauge $A_0=0$, Gauss' law implies that 
$j^0=0$ and the phase of the scalar field $\phi = {1\over\sqrt{2}}\rho 
e^{i\alpha}$ is time independent.  Furthermore, the spatial current 
vector takes on the simple form
\begin{equation}
j^i = 2e^2\rho^2A^i. \label{2.4}
\end{equation}
We thus arrive at the following system of equations for the real
scalar fields $\rho$ and the vector potential $A_i$:
\begin{eqnarray}
\ddot{\rho} + (m_0^2 + e^2A^2) \rho &=& 0, \qquad \nonumber \\
\ddot{A}_i + e^2\rho^2A_i &=& 0, \label{2.5}
\end{eqnarray}
where a dot indicates a time derivative.

Assuming that only a single component of $A_i$ is nonvanishing, this
system is classically equivalent to the well-known two-dimensional
dynamical system with quartic potential $x^2y^2$ which exhibits a
strong chaotic behavior.  The $x^2y^2$-model appears in various
contexts in science including chemistry, astronomy, astrophysics,
cosmology and most interesting for us, in the homogeneous limit of the
Yang-Mills equations \cite{MST81} (see \cite{TSB94} for details and 
references).

\subsection{Conditions for Dynamical Chaos}

There are various methods for establishing the chaoticity of the system
(\ref{2.5}).  For us most useful is the Toda-Brumer criterion
\cite{MT74,PB74}, which is based on the study of the stability matrix of 
the potential energy.  Denoting the coordinates as $q_i\;(i=1,\ldots n)$
the stability matrix for the potential $U(q)$ is
\begin{equation}
U''_{ik}(q) \equiv {\partial^2U\over \partial q_i\partial q_k}\quad 
(i,k=1,2,\ldots,n), 
\end{equation}
where we assume that the system is conservative with quadratic separable 
kinetic energy.  If the determinant of this matrix (proportional to the 
Gaussian curvature of the potential) is negative, the system is chaotic.
It is worthwhile remarking here that the use of this local
hyperbolicity criterion as a condition of global instability requires 
some caution.  However, for potentials like the $x^2y^2$ potential, the 
Toda-Brumer criterion of chaoticity is quite effective, establishing a 
necessary condition for global chaos.  Returning to our dynamical system 
(\ref{2.5}) we write the corresponding potential
\begin{equation}
U_{\rho A} = {1\over 2}m_0^2\rho^2 + {1\over 2}e^2\rho^2A^2,
\label{2.7}
\end{equation}
and the stability matrix
\begin{equation}
U_{\rho A}^{\prime\prime} = \left(
\begin{array}{cc}
m_0^2+e^2A^2 &\quad 2e^2\rho A \\
2e^2\rho A &\quad e^2\rho^2
\end{array}
\right) \label{2.8}
\end{equation}
which immediately gives the following condition for the onset of chaos:
\begin{equation}
e^2A^2 > e^2A_{cr}^2 = {m_0^2\over 3} \label{2.9}
\end{equation}
Note that $A_{cr}$ is independent of $\rho$, the amplitude of the
scalar field.  The corresponding minimal energy for the onset of chaos is
\begin{equation}
E_{cr} = {2\over 3} m_0^2\rho^2. \label{2.10}
\end{equation}
At the classical minimum of the potential $E_{cr} =0$.

We conclude that classical {\it massless} scalar electrodynamics is
strongly chaotic in the long wavelength limit for any magnitude of the 
spatially homogeneous fields.  Chaos begins at zero value of the
energy of the system.  The mass $m_0$ of the scalar field sets a threshold 
for chaos.  As we will see later, the threshold is influenced by: i) the 
self-coupling $\lambda\phi^4$ of the scalar field and ii) the quantum 
corrections to the classical potential (\ref{2.7}), both increasing the 
threshold for chaos.\footnote{The spatially uniform classical
SED with its chaoticity and the corresponding quantum mechanical
aspects leading to the suppression of chaos have been intensely
studied by the Los Alamos group \cite{CM89,CDMR}.}  Quantum corrections lead to 
the formation of a nontrivial minimum in the effective potential of SED and, 
thus, produce non-zero mass (in the case of massless classical SED) or 
increase the ``classical'' mass $m_0$.  The self-interaction of the scalar 
field also increases the threshold for chaos (for $\lambda >0)$.

\subsection{Self-interacting SED}

We begin with the study of the self-interaction of the scalar field at
the classical level.  Adding to the potential (\ref{2.7}) the quartic
self-interaction $\lambda\phi^4$ we easily obtain the following
modified conditions for the onset of chaos:
\begin{equation}
e^2A_{cr}^2 = \lambda\rho^2 + {m_0^2\over 3} \label{2.11}
\end{equation}
and
\begin{equation}
E_{cr} = {3\over 4} \lambda\rho^4 + {2\over 3}
m_0^2\rho^2. \label{2.12}
\end{equation}
At the classical minimum $\rho=0$, still $E_{cr}=0$.

For the case of the spontaneously broken gauge symmetry at the
classical level 
\begin{equation}
U_{\rho A} = {1\over 2} e^2A^2\rho^2 + {\lambda\over 4}
(\rho^2-v^2)^2.  \label{2.13}
\end{equation}
With the vacuum expectation value of the scalar field $v$ and its 
mass $m_s^2=2\lambda v^2$ we obtain the following critical 
magnitude of the gauge field potential $A_{cr}$ necessary for the onset 
of chaos:
\begin{equation}
e^2A_{cr}^2 = \lambda \left(\rho^2-{v^2\over 3}\right). \label{2.14}
\end{equation}
At the minimum of the potential (\ref{2.13}), $\rho=v$, this gives
\begin{equation}
e^2 A_{cr}^2 = {m_s^2\over 3}, \label{2.15}
\end{equation}
which coincides with the condition (\ref{2.9}) for the onset 
of chaos for the case of {\it massive} classical SED without 
self-interaction. 

Substituting $A_{cr}^2$ into (\ref{2.13}) and minimizing 
$U_{\rho A}$ with respect to $\rho$ we find:
\begin{equation}
\rho_{cr}^2 = {4\over 9}v^2 
\end{equation}
and
\begin{equation}
e^2A_{cr}^2 = {\lambda v^2\over 9} = {m_s^2\over 18}
\end{equation}
giving the minimal energy
\begin{equation}
E_{cr} = {11\over 108} \lambda v^4 \label{2.16}
\end{equation}
necessary for the onset of chaos \cite{KK89}.  Note that in the case of 
SSB the coupling constant $\lambda$ is absorbed in the definition of the 
mass generated by SSB, whereas in the corresponding case without SSB but 
with $m_0\not= 0$, $\lambda$ enters in the condition (\ref{2.11}) along 
with $m_0$.

Concluding this section, we sum up the results obtained for classical
SED:
\begin{itemize}

\item Classical massless SED is chaotic in its long wavelength limit
at any magnitude of the fields.  No threshold values exist for the
scalar and vector fields and chaos begins at zero energy of the system.

\item Introduction of a mass of the scalar field (directly or by
SSB$^2$) sets threshold values of the fields beyond which the system is 
chaotic.

\end{itemize}

It is reasonable to expect that the inclusion of a spatial dependence
of the fields, while not eliminating the general chaotic behavior of
the system under study, will result in a more complicated and rich
picture.

Finally, let us remark that spontaneously broken SED in its spatially 
homogeneous limit may serve as a model of the uniform superconductor in 
a homogeneous time dependent electrical field with its possible phase
transitions.\footnote{There also exists a close analogy between this 
phenomenon and the elimination of the chaos in SU(2) Yang-Mills model 
by the Higgs condensate \cite{SGM81}, where, depending on the value of 
the Higgs condensate, one observes a phase transition of the type 
``order--disorder''.}

\section{Quantum Corrections to Massless Scalar Electrodynamics}

\subsection{General Considerations}

It is generally believed that quantum fluctuations lead to the 
suppression of the most characteristic manifestations of dynamical 
chaos.\footnote{But not all of them: Important properties of stationary
states such as the distribution of energy level or interference patterns
(e.g., scars) reflect the intrinsic chaoticity of the corresponding 
classical systems \cite{LER92,EM84,EMP89,ZM90}.}  For the mechanical systems it is 
obvious:  the discreteness of the phase space imposed by the quantum nature 
of the system suppresses or even eliminates the long-time random behavior 
of the classically chaotic systems that is characterized by the positivity 
of the Lyapunov exponents.  Indeed, the non-stationary evolution of a 
quantum mechanical system which classically is chaotic (SED) is 
characterized by vanishing Lyapunov exponents \cite{CDMR}.

For the field theory with its infinite number of degrees of freedom
the situation is not so straightforward.  It is well established that
various field theories, among which are the spherically symmetric
Yang-Mills equations \cite{MPS88}, the Yang-Mills-Higgs equations in the
interior of a 't~Hooft-Polyakov monopole \cite{MPS89}, and the equations 
of general relativity \cite{BR,Rugh}, exhibit dynamical chaos in the 
classical limit.  Here, as in the case of the mechanical systems, the 
basic question of the competition and interference between the highly 
unstable classical fluctuations responsible for chaos and the quantum 
fluctuations of the interacting fields arises.  Do the quantum 
fluctuations suppress the chaoticity of the classical field theory?
Although practically all methods of the quantization of fields about 
chaotic classical solutions encounter the problem of instability there 
does not exist a proven way to avoid or circumvent this delicate problem 
(see, however, \cite{FGM95}).\footnote{One might be tempted to say that 
such theories should not be considered as a basis of a {\it local} quantum 
field theory, and confine oneself to integrable systems.  Not sharing
this opinion, we believe that this view reflects our inability to go
beyond the saddle-point approximation to the functional integration
by analytical means.  Furthermore, these instabilities of the 
field theory may be associated with the decay properties of
correlation functions (see below).}

In this paper, we also do not propose a general recipe for how to
quantize  a field theory that is chaotic in the classical limit, and 
confine ourselves to the loop expansion taking as a basis the chaotic 
classical theory.  Our treatment is based on the notion of the effective 
potential.  We consider here, as an example, mostly the massless SED 
since it is free from the well-known difficulties arising when, for the 
case of SSB, the new minimum lies far outside the validity of the one-loop
approximation \cite{CW73}.

We write down the effective potential $\Gamma(\phi_c)$ as the minimal 
expectation value of the Hamiltonian $\langle\psi\vert H\vert\psi\rangle$ 
in the normalized state $\vert\psi\rangle$, wherein the field $\phi(x)$ 
has a given constant expectation value $\langle\psi\vert\phi(x)
\vert\psi\rangle=\phi_c$.  Since for the calculation of the $\Gamma(\phi_c)$
one must consider the space dependence of the fields as well as their time 
dependence, the phase of the scalar field cannot be eliminated in the
$A_0=0$ gauge and, writing $\phi=\varphi_1+i\varphi_2$, two real scalar
fields $\varphi_1$ and $\varphi_2$ enter.  But the effective potential can 
depend only on $\vert\phi\vert^2=\varphi_1^2+\varphi_2^2$, so one may take
$\varphi_2=0$ and compute only graphs with $\varphi_1$-external lines 
\cite{CW73}.  The Landau gauge is most appropriate and economical here.

As it is known, the mass matrix computed from the classical potential
enters into the expression of the effective potential for two
interacting fields.  The one-loop effective potential for massless SED 
can be written as
\begin{equation}
\Gamma^{(1)}(\rho,A) = U_{\rho A}+ {\hbar\over 64\pi^2} {\rm tr}
\left[ (U_{\rho A}^{\prime\prime})^2 \ln \left( {U_{\rho
A}^{\prime\prime} / \mu^2}\right) \right] \label{3.1}
\end{equation}
where $U_{\rho A}$ is the classical potential (\ref{2.7}) and the
matrix $U''_{\rho A}$ is given by (\ref{2.8}).  Here $\mu^2$ is the 
renormalization point.

From the definition of the effective potential, it is evident that the
exact effective potential must be real.  The approximate calculation of 
this quantity in the loop expansion leads to regions of complexity 
which are impossible to eliminate for the case of the classically chaotic 
system under study.  However, as we discuss below, this does not imply 
that complexity makes the effective potential meaningless.  We see from 
(\ref{3.1}) that for a classically chaotic system characterized by 
(Toda-Brumer condition)
\begin{equation}
{\rm det} \left\{ U''_{\rho A} \right\} = {\partial^2 U\over\partial
A^2} {\partial U\over\partial\rho^2} - \left(
{\partial^2 U\over\partial A\partial\rho}\right)^2 <0  
\end{equation}
the one-loop effective potential becomes complex for almost all values
of the fields $A,\rho$ and not only for some finite range of the fields
as it occurs, e.g., in the case of SSB at the tree level for
non-chaotic systems.  Massless SED and the free Yang-Mills theory in
the limit of homogeneous fields are such systems.

It is possible to say that the complexity of the
loop-expanded\footnote{For renormalizable theories this complexity
survives at any finite order of the loop expansion.} effective potential 
is one more relic of the chaos of the initial classical theory.  The 
imaginary part of the effective potential signals not only the instability 
of the field configuration, but it is a general consequence of the chaos 
of the classical system.

From these considerations one can understand why all efforts to
eliminate the imaginary part \cite{NO78} of the one-loop effective 
potential for uniform chromomagnetic field of the pure Yang-Mills
theory in Minkowski space \cite{BMS77} were unsuccessful.  Stable 
radiative corrections in Minkowski-space require a stable classical
configuration.  But such configurations (instantons) are known only in 
the Euclidean space-time.  We postpone further considerations of this 
issue until section V, except for one related remark.  The presence of 
the imaginary part in the one-loop effective potential is intrinsically
linked to the asymptotic freedom of the non-Abelian gauge fields 
\cite{Ole81,NO78}.  It is worth noting that recently this unstable mode was 
detected directly in Monte Carlo simulations of the lattice gauge
theory\cite{LP95}.

Below, we temporarily avoid the complications caused by the imaginary
part of the potential (\ref{3.1}) considering only the effect of the 
quantum corrections along the axis $A=0$.  This ``projection'' retains the
picture of the SSB by the quantum corrections by suitable choice of 
$\mu$ in (\ref{3.1}), because the actual minima of the real part of 
$\Gamma^{(1)}$ occur on the axes $A=0$ and $\rho=0$, as a numerical
evaluation of the one-loop effective action in the $\rho-A$ plane
shows.

\subsection{One-Loop Effective Potential}

At this stage, we turn to consider the one-loop corrected effective 
potential for massless SED with self-interaction of the scalar field.
We begin with the case without SSB at the classical level.
Following \cite{CW73}, the one-loop effective potential for massless
scalar SED with self-interaction $\lambda\phi^4$ is:
\begin{eqnarray}
\Gamma^{(1)} (\rho,A;M^2) &=& {1\over 2} e^2A^2\rho^2 +
{\lambda\over 4} \rho^4 \nonumber \\
&+& {5\rho^4\over 32\pi^2} \left( \lambda^2 + 
{3e^4\over 10}\right) \left[ \ln {\rho^2\over\mu^2} - {25\over 6}\right]
\label{3.3}
\end{eqnarray}

The quantum corrections lead to a new minimum of the potential 
at $A=0$ but $\rho=\bar\rho \not= 0$ instead of $A=\rho=0$.  Implementing 
the standard procedure of dimensional transmutation we write:
\begin{equation}
\Gamma^{(1)}(\rho,A;\bar\rho) = {1\over 2} e^2A^2\rho^2 +
{5\rho^4\over 32} \left(\lambda^2 + {3e^4\over 10} \right) \left[
\ln {\rho^2\over\bar\rho^2}-{1\over 2}\right], \label{3.4}
\end{equation}
where the ${\lambda\over 4}\rho^4$ term of the classical potential is 
absorbed in the subtraction point of the logarithm.\footnote{In
contrast to \cite{CW73}, we retain $\lambda^2$ with respect to $e^4$ 
here.}

The effective potential now has a minimum value
\begin{equation}
E_0^{(1)} \equiv \Gamma^{(1)}(\rho,A;\bar\rho)\Big\vert_{{\rho=\bar\rho 
\atop A=0}} = -{5\bar\rho^4\over 64\pi^2} \left( \lambda^2 + 
{3e^4\over 10}\right), \label{3.5}
\end{equation}
which lies below the classical vacuum $E_0^{(0)} = U_{\rho
A}\vert_{\rho=A=0} = 0$.   The masses of the scalar boson and photon are
\begin{equation}
m_s^2 =
{\partial^2\Gamma^{(1)}\over\partial\rho^2}\Big\vert_{{\rho=\bar\rho
\atop A=0}}  = \delta m_{\lambda}^{(1)^2} +\delta m_e^{(1)^2}
\label{3.5a}
\end{equation}
\begin{equation}
m_A^2 = {\partial^2\Gamma^{(1)}\over\partial A^2}
\Big\vert_{{\rho=\bar\rho \atop A=0}} = e^2\bar\rho^2 \label{3.6}
\end{equation}
where
\begin{equation}
\delta m_{\lambda}^{(1)^2} = {5\lambda^2\over 4\pi^2}\bar\rho^2, \qquad
\delta m_e^{(1)^2} = {3e^4\over 8\pi^2} \bar\rho^2
\label{3.7}
\end{equation}
are the mass quantum corrections to the classically massless
scalar boson generated by the scalar self-coupling and scalar-photon
coupling, respectively, as shown in Fig. 1.

We now consider (\ref{3.4}) for the case of spatially uniform fields 
$\rho(t)$, $A(t)$ and apply the Toda-Brumer criterion for the 
onset of chaos:
\begin{equation}
{\det} \left\{ \Gamma^{(1)''} (\rho,A;\bar\rho)\right\} = 0.
\label{3.8}
\end{equation}
We obtain, from (\ref{3.8}), the critical value of $A$ beyond which
the chaos sets in:
\begin{equation}
e^2A_{cr}^2 = {5\rho^2\over 8\pi^2} \left(\lambda^2 + {3e^4\over
10}\right) \left[ \ln {\rho^2\over\bar\rho^2} + {2\over 5}\right].
\label{3.9}
\end{equation}
Taking (\ref{3.9}) in the vicinity of the new minimum $\rho=\bar\rho$, 
where our equations are reliable, we arrive at the relation 
\begin{equation}
e^2A_{cr}^2 = {m_s^2 \over 3}, \label{3.10}
\end{equation}
where $m_s$ is given by (\ref{3.5a}).  This relation must be compared with 
(\ref{2.9}) and (\ref{2.15}).  The comparison shows that quantum corrections 
generate a finite threshold for the onset of chaos in massless SED, which 
classically was chaotic for an infinitesimal amplitude of the vector
field.  Let us note here that (\ref{3.10}) also agrees with the result
obtained when one includes the self-interaction
$\lambda\phi^4$ at the classical level with $m_0\not= 0$,
inserting the classical minimum $\rho=A=0$ into (\ref{2.11}).

Let us consider next the one-loop corrections for massless SED with SSB 
at tree level.  Adding to the classical potential (\ref{2.13}) the one-loop
quantum corrections and eliminating the renormalization scale
$\mu$ by the new minimum of the scalar field $\bar\rho$, we obtain:
\begin{eqnarray}
\Gamma_{\rm SSB}^{(1)} (\rho,A;v,\bar\rho) = {1\over 2}e^2\rho^2A^2
+{\lambda\over 4}(\rho^2-v^2)^2 \nonumber \\
\qquad\qquad +{5\rho^4\over 32\pi^2} a \left[ \ln
{\rho^2\over \bar\rho^2} -b-{1\over 2}\right] 
\end{eqnarray}
with
\begin{equation}
a\equiv \lambda^2+{3e^4\over 10}, \qquad
b\equiv {8\pi^2\over 5a} \left( 1-{v^2\over\bar\rho^2} \right).
\label{??}
\end{equation}
Following the same steps as above, we obtain the critical values of
$A$ for the onset of chaos:
\begin{equation}
e^2A_{cr}^2 = \lambda \left( \rho^2 - {v^2\over 3} \right) + {5a\over
8\pi^2} \rho^2 \left[ \ln {\rho^2\over\bar\rho^2} - b +{2\over
3}\right]. \label{???}
\end{equation}
Minimizing $\Gamma_{\rm SSB}^{(1)}(\rho,A_{cr};v,\bar\rho)$ with
respect to $\rho$ and inserting the tree level values for $\bar\rho$
and $\rho_{cr}$ into the terms describing the quantum corrections, we
get:
\begin{equation}
\lambda\rho_{cr}^2 \approx {4\over 9}\lambda v^2 +{0.05\over 6\pi^2}
\left( {\lambda^2+3e^4\over 10} \right) v^2. \label{????}
\end{equation}
Finally, we arrive at the minimal energy for the onset of chaos:
\begin{equation}
E_{cr} \approx {11\over 108} \lambda v^4 - {4\over
81\pi^2} \left( \lambda^2 + {3e^4\over 10}\right) v^4. \label{unknown}
\end{equation}
The sign of the second term in (\ref{unknown}) is not surprising since
the shift of the vacuum energy for the case without SSB (see (\ref{3.5})) 
is bigger than for the case with SSB.

Finally, a slightly philosophical remark.  From the above results one
may say that massless charged scalar particles do not exist in nature
due to the unavoidable quantum fluctuations which give them mass,
eliminating the chaotic behavior of the field theory in the presence
of an infinitesimal electromagnetic field.

\subsection{Two-Loop Corrections}

The situation is repeated for the case of the two-loop effective
potential, which we describe briefly.  After minimization we may write 
the effective potential $\Gamma^{(2)}$ \cite{Wein93} up to
order $e^6$ (here we discard $\lambda^2$ against $e^4$):
\begin{eqnarray}
\Gamma^{(2)}(\rho,A;\tilde\rho) &=& {1\over 2} e^2A^2\rho^2 \nonumber \\
&+& {3e^4\over 64\pi^2} (1+Ce^2)\rho^4 
\left( \ln {\rho^2\over\tilde\rho^2}-{1\over 2} \right) \nonumber \\
&+& {5e^2\over 512\pi^4} \rho^4\ln^2
\left[{\rho^2\over\tilde\rho^2}\right] \label{3.11}
\end{eqnarray}
where $\rho=\tilde\rho,\; A=0$ is the minimum of the potential. The 
constant $C$ (of order of unity) is defined by the precise specification 
of the renormalization conditions.

Figure 2 shows the two-loop graph which gives the $O(e^6)$ contribution 
in Landau gauge.  Purely scalar loops enter at the higher order $O(e^8)$.  
Again, as in (\ref{3.4}), the classical self-interaction is absorbed in 
the process of the dimensional transmutation.  The minimal value of the 
two-loop effective potential is
\begin{equation}
\Gamma^{(2)}\bigg\vert_{{\rho=\tilde\rho \atop A=0}} = - {3e^4\over
128\pi^2}\; (1+ Ce^2)\tilde\rho^2. \label{3.15}
\end{equation}
As for the case of the one-loop correction, we obtain the expression
for the critical value of $A$ beyond which the system is chaotic (at
$\rho=\tilde\rho)$:
\begin{equation}
3e^2A_{cr}^2 = {3e^4\tilde\rho^2\over 8\pi^2} +
{15e^6\tilde\rho^2\over 128\pi^4} \left[ {2\over 3} + {16\pi^2\over 5}
C\right]. \label{3.16}
\end{equation}
(\ref{3.3}) can be written in the form analogous to (\ref{3.10})
\begin{equation}
e^2A_{cr}^2 = {1\over 3} \left( \delta m_e^{(1)^2} + 
\delta m_e^{(2)^2}\right) = {m_s^2\over 3}, \label{3.17}
\end{equation}
where $\delta m_e^{(2)^2}$ is the two-loop scalar boson mass correction
up to $O(e^6)$ in the limit $\lambda^2 \ll e^4$.  The condition (\ref{3.17}), 
as well as (\ref{3.10}), can be embodied in the general relation
\begin{equation}
3e^2A_{cr}^2 =
{\partial^2\Gamma\over\partial\rho^2}\bigg\vert_{{\rho=\rho_{\rm min}
\atop A=0\hfill}}, \label{3.15x}
\end{equation}
which is a consequence of the specific structure of SED and the 
$x^2y^2$-model.

\section{On the Imaginary Part of the One-Loop Effective Potential}

Up to now we have avoided the complications arising from the
imaginary part of the effective potential---appearing due to the
chaoticity of classical SED---by ``projecting'' the quantum
corrections onto the axis $A=0$.  The practical argument justifying such 
a limitation and giving a transparent physical interpretation of 
the imaginary part of $\Gamma^{(1)}$ was formulated in \cite{EJW87}.  
Weinberg and Wu modified the definition of the effective potential by 
adding the further restriction that the wave functional for the state 
$\vert\psi\rangle$, which minimizes the Hamiltonian subject to the 
condition $\langle\psi\vert\phi(x)\vert\psi\rangle =\phi_c$, be 
concentrated on configurations $\phi(x)$ about $\phi_c$.  Then, in the 
leading order, the redefined potential $\tilde\Gamma^{(1)}(\phi_c)$ 
will differ from the classical one, $U(\phi_c)$, by the zero-point 
energies of the quantum fluctuations $\phi(x)$ about $\phi_c$ and will 
thus agree with the real part of the one-loop effective potential.

But what happens if the region $\phi(x)\approx \phi_c$ is unstable 
(chaotic) as expressed by the condition det$\{U''_{\rho A}\}<0$?  Then 
the effective potential acquires an imaginary part which describes the 
decay rate per unit volume, or the damping rate of some correlation 
functions (see \cite{EJW87,Wein93} for details).  This interpretation of the 
imaginary part of $\Gamma^{(1)}$ allows one to understand the connection 
between the chaoticity of the underlying classical systems expressed by 
the positivity of the Lyapunov exponents and the damping (decay) 
properties of the corresponding quantum system.

Following \cite{EJW87} we consider the illuminating example of the 
upside-down harmonic oscillator with the potential $-{1\over 2}
\eta^2q^2$.  Of course, there is no lower bound on the energy, but the 
requirement that fluctuations be confined to small amplitudes, thus 
considering states with wave functions concentrated about small values 
of $q$: $\langle\psi\vert q^2\vert\psi\rangle \le {a\over\eta}$, puts a 
lower bound on the Hamiltonian.  The ``decay rate'' at large time is given 
by 
\begin{equation}
\vert\langle\psi(0)\vert\psi(x)\rangle\vert^2 \sim e^{-\eta t},
\label{whoknows}
\end{equation}
whereas the ``quantum analogue'' of the Lyapunov exponent---spreading of 
the state $\psi$ in time---is
\begin{equation}
[\langle\psi(x)\vert q^2\vert\psi(x)
\rangle]^{1/2} \sim e^{\eta t}. \label{whocares}
\end{equation}
Thus, it is not surprising that the relation between the positive
maximal Lyapunov exponents $\lambda_0$ characterizing classical chaos
of the SU(2) and SU(3) gauge theories (without quarks) and the 
corresponding analytically calculated damping rates $\gamma$ in the 
thermalized system of gauge bosons $\lambda_0=\gamma$ is found
numerically on the lattice \cite{BGMT94,CG93}.\footnote{The factor 2 
which appears in this relation in \cite{BGMT94,CG93} is a result of the 
specific parametrization of the plasmon pole leading to the decay of the 
energy density as exp $(-2\gamma t)$ \cite{BM95}.}

\section{Instability of the Effective Potential for Non-Abelian Gauge
Theories}

We repeatedly emphasized in this paper the relation between the
instability of the (one-loop) effective potential and the intrinsic
chaoticity of the classical non-Abelian gauge theories in Minkowski 
space.  To avoid this kind of instability, one needs to start from 
a stable classical configuration.  Pure gauge theories (without
fermions, thermal effects, or Higgs fields) do not possess such stable 
classical states in Minkowski space (see \cite{TSB94} for details on the 
sources of the stabilization of classical gluon fields). 
The problem in pure gauge theories is that it is not possible to
generate a dynamical mass for the gauge bosons perturbatively, in
contrast to theories involving scalar fields.  There exists, of
course, a non-perturbative mechanism of mass generation:
confinement.  This property of non-Abelian gauge theories avoids the
existence of propagating massless modes of the gauge field; only
color-singlet bound states exist as physical fluctuations around the 
vacuum state.

It is widely believed that confinement in non-Abelian gauge theories
is due to the presence of a vacuum condensate of chromomagnetic
monopoles.\footnote{This mechanism has been rigorously established for
confinement in supersymmetric non-Abelian gauge theories \cite{SW94}.}  
A complete description of this mechanism is clearly beyond the reach 
of perturbation theory, because no classically stable magnetic monopole 
solutions exist in pure gauge theories.  However, we will demonstrate 
below that the condition of the stability of the one-loop effective 
potential allows us to derive a lower limit on the density of magnetic 
monopoles in the gauge field vacuum.

It is well-known that the one-loop effective potential for a uniform
chromomagnetic field \cite{BMS77} is unstable \cite{NO78}.  From our 
point of view, developed in the present paper, it is the result of the 
chaoticity of the corresponding classical Yang-Mills theory.  This raises 
the question, which mechanism can induce the stability of the effective
potential in the infrared domain?  The natural approach is to consider 
uniform non-Abelian flux tube configurations.  The properties of various 
types of such tubes were studied at the classical level \cite{NO78,GOL93}.  
It was shown that the stability can be  achieved only by confining the 
chromomagnetic flux, not only in the transverse direction, but also along 
the direction of the chromomagnetic field.  This condition can be naturally 
realized if one ends the flux on sources of the magnetic field lines.  
These sources can be interpreted as magnetic monopoles.

For the stability of such configurations the length $L$ must be smaller 
than $L_0 \approx \pi/\sqrt{gH}$ to eliminate small momenta $k < 
\sqrt{gH}$ contributing to the instability.  We can utilize this
condition to estimate the density of magnetic monopoles in the QCD
vacuum.  The real part of the renormalized one-loop effective potential 
for a constant chromomagnetic field $H$ is \cite{BMS77}
\begin{equation}
\Gamma^{(1)}(H) = {1\over 2}H^2 + {11N_c\over 96\pi^2} (gH)^2 \left[
\ln {gH\over\mu^2}-{1\over 2}\right], \label{48}
\end{equation}
where $N_c$ is the number of colors and $\mu^2$ is the renormalization
point.  The minimum of the effective potential,
\begin{equation}
gH_0 = \mu^2 \exp \left( -{48\pi^2\over 11N_cg^2(\mu)} \right), \label{49}
\end{equation}
is a renormalization group invariant.

Now consider the corresponding magnetic flux in a region of
length $L$.  Gauss' law states that the magnetic monopole charge
density (per area) terminating the magnetic flux lines equals
$H_0$.  Because the elementary magnetic monopole charge in
non-Abelian gauge theories is $4\pi/g$ \cite{GtH74}, the required density of
magnetic monopoles is
\begin{equation}
n_M = 2 {gH_0\over 4\pi L}, \label{50}
\end{equation}
where the factor 2 counts monopoles and antimonopoles.  Inserting the
condition $L\le L_0$ for the vacuum stability, we obtain
\begin{equation}
n_M \ge {gH_0\over 2\pi L_0} = {(gH_0)^{3/2} \over 2\pi^2}. \label{51}
\end{equation}

We can obtain a numerical estimate for the monopole density, if we
identify $gH_0$ with the gluon condensate obtained in the QCD sum rule
approach \cite{SVZ79}
\begin{equation}
(gH_0)^2 = \langle g^2 F_{\mu\nu}^aF^{a\mu\nu}\rangle \approx 0.5\, {\rm
GeV}^4. \label{new}
\end{equation}
This yields a value for a monopole density:
\begin{equation}n_M \approx 0.03\, {\rm GeV}^3 \approx 4\, {\rm fm}^{-3}.
\label{density}
\end{equation}

Of course, the reliability of the one-loop approximation is
questionable near the minimum of the effective potential, because the
vacuum field $H_0$ is not strong.  However, it is noteworthy that
this very simple consideration provides a relation between the
strength of the monopole condensate and the strength of the gluon
condensate in QCD, which can be tested by lattice calculations.

One can further develop this essentially perturbative picture of
the chromomagnetic condensate, based on (\ref{48}), by the extension
of the electromagnetic duality principle to the weak-strong coupling
duality \cite{SW94,MO77} connecting the weakly coupled chromomagnetic
phase with the strong coupling regime of the chromoelectric phase with
its color confinement.  This picture is in accord with Monte Carlo
simulations \cite{LP95} where the unstable modes in a chromomagnetic
background were observed.  These results support the idea that the QCD
vacuum behaves in the continuum limit as a quasi-Abelian magnetic
condensate with the properties of a dual superconductor.

\section{Conclusions}

What lessons can we draw from the results presented here?  First of
all, we have shown that the onset of chaoticity of the classical
fields in theories such as SED is delayed by the radiative
corrections.  In the case of massless SED, which is chaotic for all
energies at the tree level, the radiative corrections introduce a
threshold for the onset of chaos.

The classical chaoticity, in turn, leads to the instability of the
corresponding effective potential, presumably, at any finite order of
the loop expansion.  This interdependence may explain the failure of
numerous attempts to eliminate the Nielsen-Olesen instability
\cite{NO78} of the one-loop effective potential for a uniform
chromomagnetic field in the pure Yang-Mills theory \cite{BMS77}.

Since the true effective potential is known to be always a real and
convex function of the field expectation values, the instabilities
associated with deterministic chaos must be absent in the full quantum
theory.  Higher-order (non-Gaussian) quantum fluctuations provide the
mechanism for this phenomenon.  We have already discussed the
double-well oscillator where the ground state can be approximated as
the sum of {\it two} Gaussians.

The suppression of chaoticity can sometimes also be seen at the
classical level, e.g., if one includes the anharmonicity stabilizing
the inverted oscillator.  Consider the equation
\begin{equation}
\ddot\varphi - m^2\varphi +\lambda\varphi^3 = 0.
\end{equation}
In the absence of the anharmonic term, the equation has a solution
$\varphi (t)\sim e^{mt}$ for zero energy, indicating exponential
instability.  If we include the $\lambda\varphi^3$ term, the
zero-energy solution becomes \cite{CT92}
\begin{equation}
\varphi(t) = \sqrt{{8m^2\over\lambda}} {e^{\pm m(t-t_0)}\over 1+
e^{\pm 2m(t-t_0)}}, 
\end{equation}
which behaves as $e^{-mt}$ for large times.  Unfortunately, it is not
known how to perform functional integrals in quantum field theory
beyond the Gaussian approximation by analytical techniques.

This raises the question whether, in a given theory, it is possible to
find a stable classical configuration in Minkowski space around which
the theory can be quantized.  Several mechanisms are known \cite{TSB94}
which generate stable solutions (and hence eliminate chaos at low
energies) in gauge theories:  mass generation by the Higgs mechanism
or topological effects, mass generation by medium polarization at
finite temperature, and stabilization of fluctuations by external
charges\cite{AAB83,LYF96}.  Although none of these mechanisms directly 
applies to the QCD vacuum, the quark vacuum condensate may have a 
stabilizing effect.  On a more general scope, the question of the possible
stabilizing role of fermions in supersymmetric Yang-Mills
theories arises.  We hope to return to these issues in the future.

\subsection*{Acknowledgements:}

This work was supported in part by a grant (DE-FG02-96ER40945) from
the U.S. Department of Energy.  One of us (SGM) thanks the Los Alamos
Theory Division for their warm hospitality during his visit, where
part of the present work was developed.

\begin{figure}
\def\epsfsize#1#2{1.3#1}
\centerline{\epsfbox{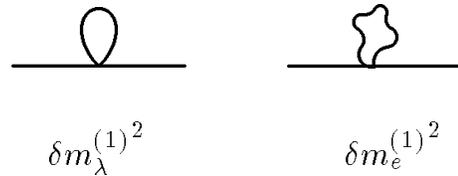}}
\caption{One-loop contributions to the scalar boson mass.  Solid
lines--scalar boson, wavy lines--photon.}
\end{figure}

\begin{figure}
\def\epsfsize#1#2{1.3#1}
\centerline{\epsfbox{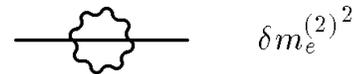}}
\caption{The two-loop graph giving $O(e^6)$ contribution to $\delta
m_e^{(2)}$.}
\end{figure}


\begin{references}

\bibitem{KSy70} K. Symanzik, {\sl Communications Math. Phys. {\bf
16}}, 48 (1970).

\bibitem{JI75} J. Iliopoulus, C. Itzykson, and A. Martin, {\sl Rev.
Mod. Phys. {\bf 47}}, 165 (1975).

\bibitem{CT72} C. Thomson, {\sl Mathematics of Statistical Mechanics},
Princeton Univ. Press, 1972; L.S. Brown, {\sl Quantum Field Theory},
Cambridge University Press, 1994, Chap. 6.


\bibitem{JSL69} J.S. Langer, {\sl Ann. Phys. (N.Y.) {\bf 41}}, 108
(1967); {\bf 54}, 258 (1969).

\bibitem{LOR76} L. O'Raifeartaigh and G. Parravicini, {\sl Nucl. Phys. {\bf
B111}}, 501 (1976).

\bibitem{YF83} Y. Fujimoto, L. O'Raifeartaigh, and G. Parravicini,
{\sl Nucl. Phys. {\bf B212}}, 268 (1983).

\bibitem{CMB83} C.M. Bender and F. Cooper, {\sl Nucl. Phys. {\bf
B224}}, 403 (1983); F. Cooper and B. Freedman, {\sl Nucl. Phys. {\bf
B239}}, 459 (1984).

\bibitem{LOR86} L.O'Raifeartaigh, A. Wipf, and H. Yoneyama, {\sl
Nucl. Phys. {\bf B271}}, 653 (1986).

\bibitem{SC74} S. Coleman, R. Jackiw, and H.D. Politzer, {\sl Phys.
Rev. {\bf D10}}, 2491 (1974); S. Coleman, Secret-Symmetry, in {\sl
Laws of Hadronic Matter, ed. A. Zichichi}, Academic Press, NY and
London, 1975.

\bibitem{EJW87} E.J. Weinberg and A. Wu, {\sl Phys. Rev. {\bf D36}}, 2474
(1987).

\bibitem{MT74} M. Toda, {\sl Phys. Lett. {\bf A48}}, 335 (1974).

\bibitem{PB74} P. Brumer, {\sl Journ. of Computational Physics {\bf
14}}, 391 (1974).

\bibitem{SGM81} S.G. Matinyan, G.K. Savvidy, and N.G.
Ter-Arutyunyan-Savvidy, {\sl Sov. Phys. JETP Letters {\bf 34}}, 590
(1981).

\bibitem{MST81} S.G. Matinyan, G.K. Savvidy, and N.G.
Ter-Arutyunyan-Savvidy, {\sl Sov. Phys. JETP {\bf 53}}, 421 (1981);
S.G. Matinyan, {\sl Sov. J. of Part. and Nucl. {\bf 16}}, 226 (1985).

\bibitem{TSB94} T.S. Bir\'o, S.G. Matinyan, and B. M\"uller, {\sl
Chaos and Gauge Field Theory}, World Scientific, 1994.

\bibitem{CM89} F. Cooper and E. Mottola, {\sl Phys. Rev. {\bf D40}},
459 (1989); Y. Kluger, J. Eisenberg, B. Svetitsky, F. Cooper, and E.
Mottola, {\sl Phys. Rev. Lett. {\bf 67}}, 2427 (1991); F. Cooper, S.
Habib, Y. Kluger, E. Mottola, J.P. Paz, and P.R. Anderson, {\sl Phys.
Rev. {\bf D50}}, 2848 (1994); F. Cooper, J. Dawson, D. Meredith, and
M. Shepard, {\sl Phys. Rev. Lett. {\bf 72}}, 1337 (1994); F. Cooper,
J. Dawson, S. Habib, Y. Kluger, D. Meredith, and M. Shepard, {\sl
Physica {\bf D83}}, 74 (1995).

\bibitem{KK89} C. Kumar and A. Khare, {\sl J. of Phys. {\bf A22}},
L849 (1989).

\bibitem{LER92} L.E. Reichl, {\sl The Transition to Chaos:  In
Conservative Classical Systems: Quantum Manifestations},
Springer-Verlag, New York, Inc., 1992.

\bibitem{EM84} E. Heller, {\sl Phys. Rev. Lett. {\bf 53}}, 1515 (1984).

\bibitem{EMP89} B. Eckhardt, G. Hose, and E. Pollak, {\sl Phys. Rev.
{\bf 39A}}, 3776 (1989).

\bibitem{ZM90} J. Zakrzewski and R. Marcinek, {\sl Phys. Rev. {\bf
42A}}, 7172 (1990).

\bibitem{CDMR} F. Cooper, J. Dawson, S. Habib, and R.D. Ryne, {\sl
Phys. Rev. D} (in press), preprint $\langle$ quant-ph/9610013$\rangle$.

\bibitem{MPS88} S.G. Matinyan, E.B. Prokhorenko, and G.K. Savvidy,
{\sl Nucl. Phys. {\bf B298}}, 414 (1988).

\bibitem{MPS89} S.G. Matinyan, E.B. Prokhorenko, and G.K. Savvidy,
{\sl Sov. J. Nucl. Phys. {\bf 50}}, 178 (1989).

\bibitem{BR} J.D. Barrow, {\sl Phys. Rev. Lett. {\bf 46}}, 963 (1981);
{\sl Physics Reports {\bf 85}}, 1 (1982).

\bibitem{Rugh} S. Rugh, in: {\sl Deterministic Chaos in General
Relativity}, NATO-ARW Proceedings, July 1993, Canada (Plenum Press,
New York, 1994) p. 359.

\bibitem{FGM95} H.M. Fried, Y. Gabelini, and B.H.J. McKellar, {\sl
Phys. Rev. Lett. {\bf 74}}, 4373 (1995); H.M. Fried and Y. Gabelini,
{\sl Phys. Rev. {\bf D51}}, 890 (1995); Y.A. Dabagyan, {\sl Phys. Rev.
Lett. {\bf 77}}, 2666 (1996).

\bibitem{CW73} S. Coleman and E. Weinberg, {\sl Phys. Rev. {\bf D7}},
1888 (1973).

\bibitem{NO78} N.K. Nielsen and P. Olesen, {\sl Nucl. Phys. {\bf
B144}}, 376 (1978).

\bibitem{BMS77} I.A. Batalin, S.G. Matinyan, and G. Savvidy, {\sl Sov.
J. Nucl. Phys. {\bf 26}}, 214 (1977); S.G. Matinyan and G.K. Savvidy,
{\sl Nucl. Phys. {\bf B134}}, 539 (1978); G.K. Savvidy, {\sl Phys.
Lett. {\bf B71}}, 133 (1977).

\bibitem{Ole81} R.J. Hughes, {\sl Nucl. Phys. {\bf B186}}, 376 (1981);
P. Olesen, {\sl Physica Scripta {\bf 23}}, 1000 (1981); N.K. Nielsen,
{\sl Am. J. Phys. {\bf 49}}, 1171 (1981).

\bibitem{LP95} A.R. Levi and J. Polonyi, {\sl Phys. Lett. {\bf B357}},
186 (1995); P. Cea and L. Cosmai, preprint
$\langle$hep-lat/9610028$\rangle$.

\bibitem{Wein93} E. Weinberg, {\sl Phys. Rev. {\bf D47}}, 4614 (1993).

\bibitem{BGMT94} T.S. Bir\'o, C. Gong, B. M\"uller, and A. Trayanov,
{\sl Int. J. Mod. Phys. {\bf C5}}, 113 (1994).

\bibitem{CG93} T.S. Bir\'o, C. Gong, and B. M\"uller {\sl Phys. Rev. 
{\bf D52}}, 1260 (1995).

\bibitem{BM95} B. M\"uller, in Proceedings of the Workshop on {\sl
Quantum Infrared Physics}, Edited by H.M. Fried and B. M\"uller (World
Scientific, 1995), p. 509.

\bibitem{SW94} N. Seiberg and E. Witten, {\sl Nucl. Phys. {\bf B426}},
19 (1994); {\bf B430}, 485 (1994).

\bibitem{GOL93} E.I. Guendelman, D.A. Owen, and A. Leonidov, {\sl Int.
J. Mod. Phys. {\bf A8}}, 4754 (1993).

\bibitem{GtH74} G. 't Hooft, {\sl Nucl Phys. {\bf B79}}, 276 (1974);
A.M. Polyakov, {\sl JETP Lett. {\bf 20}}, 194 (1974).

\bibitem{SVZ79} M. Shifman, A. Vainstein, and V. Zakharov, {\sl Nucl. Phys.
{\bf B147}}, 385, 188, 519 (1979).

\bibitem{MO77} C. Montonen and D. Olive, {\sl Phys. Lett. {\bf B72}},
117 (1977); P. Goddard, J. Nuyts, and D. Olive, {\sl Nucl. Phys. {\bf
B125}}, 1 (1977).

\bibitem{CT92} J.M. Cornwall and G. Tiktopoulos, {\sl Phys. Rev. {\bf
D45}}, 2105 (1992);  L.S. Brown, ibid {\bf 46} R4125 (1992).

\bibitem{AAB83} A. Avakyan, S. Arutyunyan, and G. Baseyan, Preprint of
Yerevan Physics Institute, Yerevan, 1983 EFI-641(31)-83
(unpublished).

\bibitem{LYF96} A.G. Lavkin, {\sl Physics of Atomic Nuclei, {\bf 59}}, 
898 (1996); T.S. Bir\'o, preprint $\langle$hep-ph/9511354$\rangle$, 1995.





\end{references}
\end{document}